\documentclass[acmtog, screen,
]{acmart}
\usepackage{dblfloatfix}
\usepackage{amsmath}
\usepackage{graphicx}
\usepackage{booktabs}
\usepackage{makecell}

\copyrightyear{2024}
\acmYear{2024}
\setcopyright{rightsretained}
\acmConference[SA Conference Papers '24]{SIGGRAPH Asia 2024 Conference Papers}{December 3--6, 2024}{Tokyo, Japan}
\acmBooktitle{SIGGRAPH Asia 2024 Conference Papers (SA Conference Papers '24), December 3--6, 2024, Tokyo, Japan}\acmDOI{10.1145/3680528.3687679}
\acmISBN{979-8-4007-1131-2/24/12}

\begin{document}
\title[Sketching With Your Voice: ``Non-Phonorealistic'' Rendering of Sounds via Vocal Imitation]{Sketching With Your Voice: \\ ``Non-Phonorealistic'' Rendering of Sounds via Vocal Imitation}

\author{Matthew Caren}
\email{mcaren@mit.edu}
\orcid{0009-0002-5061-4599}
\authornote{Equal contribution.}
\author{Kartik Chandra}
\email{kach@mit.edu}
\orcid{0000-0002-1835-3707}
\authornotemark[1]
\affiliation{
  \institution{MIT CSAIL}
  \city{Cambridge}\state{MA}\country{USA}
}
\author{Joshua B. Tenenbaum}
\affiliation{
  \institution{MIT Department of Brain \& Cognitive Sciences}
  \city{Cambridge}\state{MA}\country{USA}
}
\email{jbt@mit.edu}
\orcid{0000-0002-1925-2035}
\author{Jonathan Ragan-Kelley}
\affiliation{
  \institution{MIT CSAIL}
  \city{Cambridge}\state{MA}\country{USA}
}
\email{jrk@mit.edu}
\orcid{0000-0001-6243-9543}
\author{Karima Ma}
\email{karima@mit.edu}
\orcid{0000-0003-4180-6433}
\authornotemark[1]
\affiliation{
  \institution{MIT CSAIL}
  \city{Cambridge}\state{MA}\country{USA}
}

\begin{abstract}
We present a method for automatically producing human-like vocal \mbox{imitations} of sounds: the equivalent of ``sketching,'' but for auditory rather than visual representation. Starting with a simulated model of the human vocal tract, we first try generating vocal imitations by tuning the model's control parameters to make the synthesized vocalization match the target sound in terms of perceptually-salient auditory features. Then, to better match human intuitions, we apply a cognitive theory of communication to take into account how human speakers reason strategically about their listeners. Finally, we show through several experiments and user studies that when we add this type of communicative reasoning to our method, it aligns with human intuitions better than matching auditory features alone does. This observation has broad implications for the study of depiction in computer~graphics.
\end{abstract}

\begin{teaserfigure}
  \includegraphics[width=\linewidth]{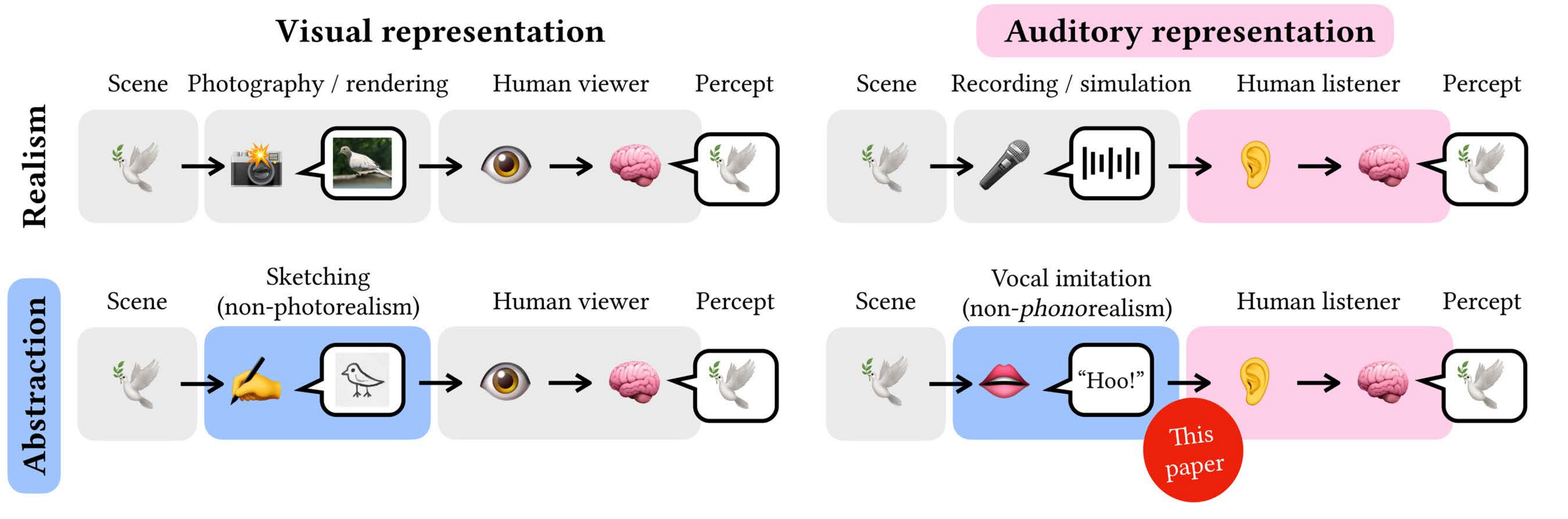}
  \caption{This paper seeks to start filling the ``missing quadrant'' of representation in graphics. Graphics has long studied how people abstract visual experience in order to create non-photorealistic sketches that induce the desired percept in the viewer's mind. In this paper, we apply the same ideas to \emph{sound}, modeling how people abstract \emph{auditory} experience to create non-\emph{phono}realistic vocal imitations that induce desired percepts in the listener's mind.
  }
  \label{fig:teaser}
\end{teaserfigure}



\maketitle

\section{Introduction}

Imagine one morning you wake up to the sound of a beautiful bird-song that you have never heard before. You cannot see the bird, but you would like very much to know its species.
Later that day, you visit an ornithologist for help. But then you have a problem: how do you communicate the bird song you heard? How can you reproduce an auditory experience in a listener's mind?

Bird songs are notoriously difficult to communicate with adjectives --- ``essentially indescribable,'' wrote \citet{hunt1923phonetics}.
Yet people, even non-specialists, nonetheless manage to informally communicate bird songs to one another all the time. To do so, they reach for a simple and intuitive mode of communication: \textbf{vocal imitation}. Without specialized training, people have the remarkable ability to use their voices to produce utterances that convey the ``essence'' of a sound to a listener: whether it is the whine of an engine to an auto mechanic, the timbre of a trombone to a musician, or the song of a bird to an ornithologist. We encourage readers to try vocally imitating: (a)~a crow cawing, (b)~an ambulance siren, and (c)~a struck~bell.

\begin{figure*}[b]
\includegraphics[width=0.33\linewidth]{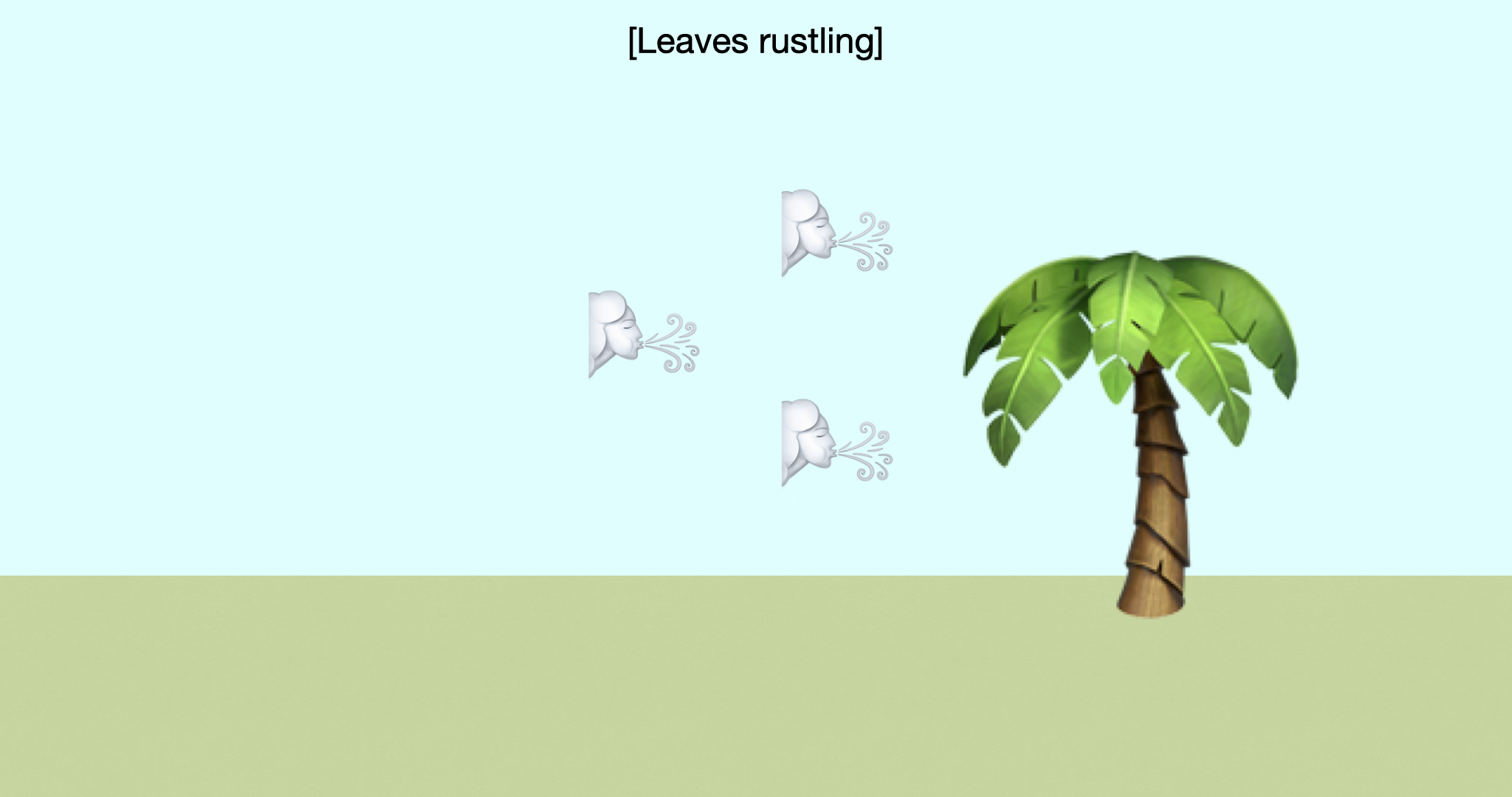}~~~
\includegraphics[width=0.33\linewidth]{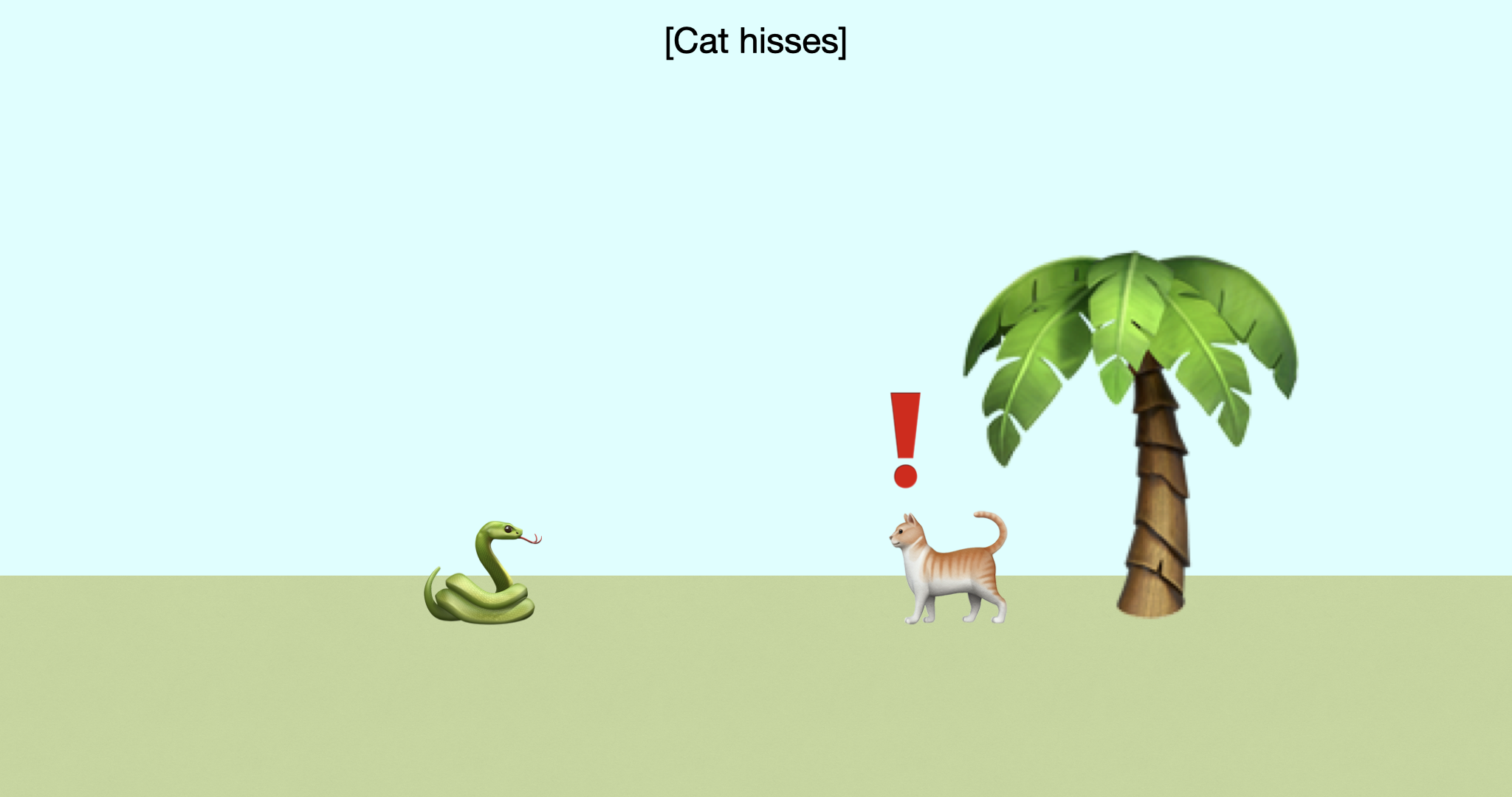}~~~
    \includegraphics[width=0.33\linewidth]{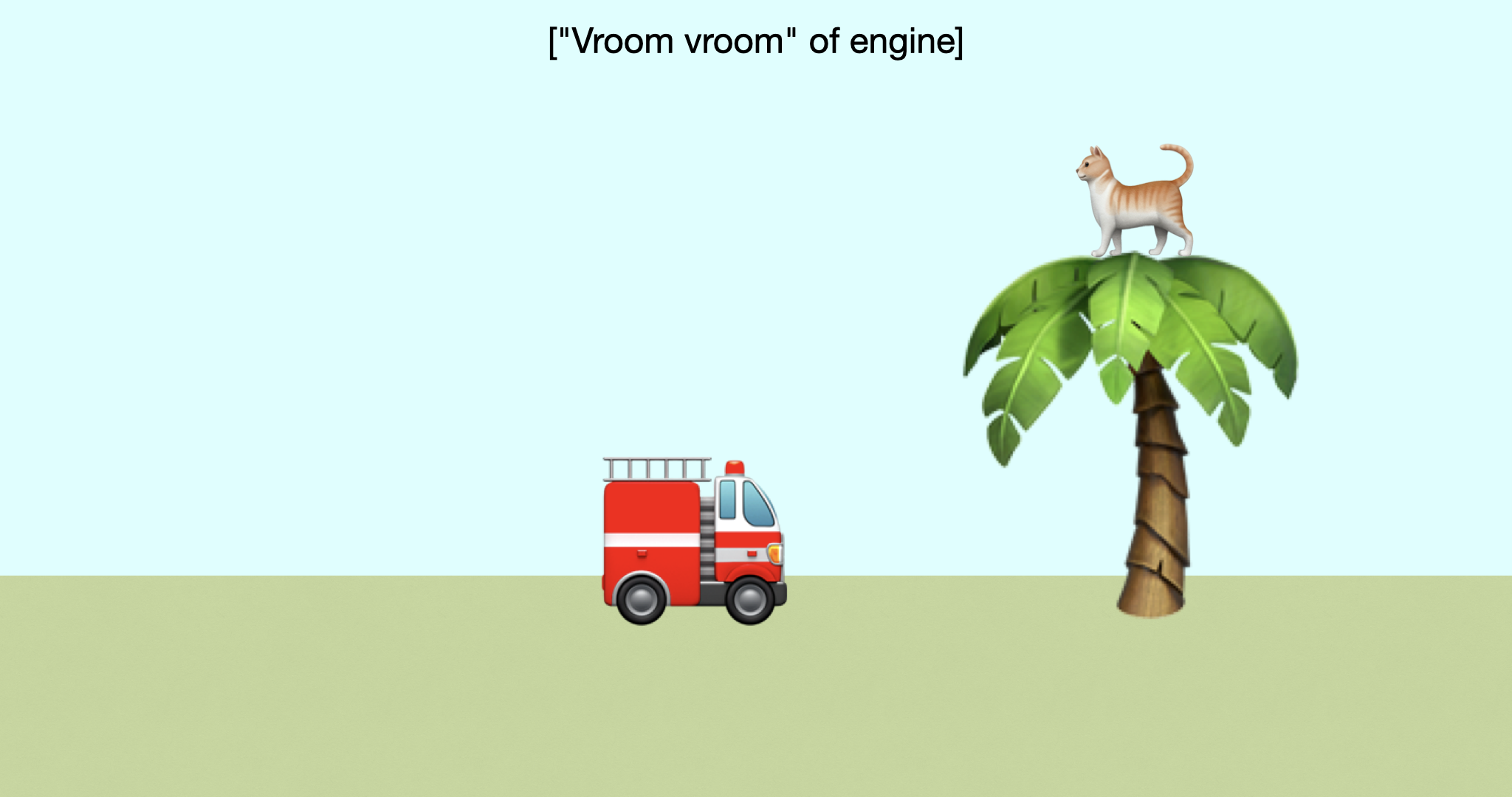}%
    \caption{Stills from a short animation, where all 11 sound effects were vocal imitations produced by our system (see supplemental materials).}
    \label{fig:vocal-foley}
\end{figure*}

In this paper, we design a method for automatically producing and understanding human-like vocal imitations. To do so, we draw an analogy between vocal imitation and \emph{sketching}.
Just like how people can quickly communicate a vast array of \emph{visual} experiences by sketching (despite the limited affordances of line drawings), people can quickly communicate a vast array of \emph{auditory} experiences by vocal imitation (despite the limited affordances of their vocal tract).

Why is this problem important to graphics?
\textbf{Our broader goal is to begin filling in the ``missing quadrant'' of representational modes in graphics} (Figure~\ref{fig:teaser}). The field has extensively studied methods for realistic visual representation (e.g.\ physically-based rendering), non-photorealistic visual representation (e.g.\ line drawing, hatching), and realistic auditory representation (e.g.\ physically-based simulation). Our work on vocal imitation begins to address ``non-\emph{phono}realistic'' auditory representation. Just as early work on sketching ultimately led to a variety of creative methods for non-photorealistic rendering of 3D~models and intuitive sketch-based interfaces for 3D~modeling (see Section~\ref{sec:background}), we hope our work here ultimately leads to a variety of creative methods for non-phonorealistic sound design, as well as new intuitive interfaces for sound designers.

For this reason, our goal is not only to \emph{produce} human-like vocal imitations: it is also to better understand the \emph{process} of vocal imitation, and gain insight into the nature of auditory abstraction. Thus, we designed our method from first principles, starting with what we know about the human vocal tract, human auditory perception, and human communication. Importantly, we do not train a model on any dataset of human vocal imitations --- nonetheless, without ever having ``heard'' a human vocal imitation, our method is able to predict human behavior, and flexibly perform a variety of additional tasks as well. This suggests that our method truly does begin to capture the essence of human vocal imitation. As such, our work contributes to the graphics community's longstanding goal of developing a computational theory of \emph{depiction} \citep{durand2002invitation}, long regarded as a key open problem in graphics \citep{foley2000getting}.

\medskip

As a first attempt at building a vocal imitation engine (Section~\ref{sec:naive}), we start with a simple controllable model of the human vocal tract (similar to the ``Pink Trombone'' \citep{pinktrombone}) that can synthesize a variety of human-like utterances. Given a target sound, we optimize the vocal tract's controls in order to make the synthesized utterance match that target sound in terms of low-level auditory features.

However, we find that this approach does not work very well: the synthesized utterances rarely match the utterances real humans produce. This should not be surprising: just like how an effective line drawing might look very different from a photograph of the object being drawn, an effective vocal imitation might sound very different from the target sound being imitated. In Section~\ref{sec:rsa}, we discuss the cognitive science basis of this phenomenon. Drawing on recent cognitive modeling techniques, we equip our method with a \emph{communicative reasoning module} that addresses this issue.

The resulting method does indeed produce human-like vocal imitations for a wide variety of sounds. For example, we produced a short animation where all sound effects were vocal imitations synthesized by our system (Figure~\ref{fig:vocal-foley}). To quantitatively evaluate our method, we compare its outputs to experimentally-collected human-generated utterances, finding a remarkably tight correlation (Section~\ref{sec:res-humans}). We also present a user study showing that when people are asked which vocal imitation they prefer for a given sound, they choose our method over the feature-matching baseline (Section~\ref{sec:res-users}). People even choose our method's utterances over \emph{human-generated} utterances 25\% of the time, and around 50\% of the time (i.e.\ chance-level parity) for several individual sounds.

Because of the way our method is designed, it can adapt flexibly to novel constraints such as restricting the speaker to whisper (Section~\ref{sec:res-whisper}). It can even be \emph{inverted} in order to \emph{understand} human-generated vocal imitations (Section~\ref{sec:res-retrieval}). This could help artists rapidly search databases of sounds that are difficult to describe in a textual query or prompt, by instead querying by vocal imitation.

\section{Background \& Related Work}\label{sec:background}

The computer graphics community has a long tradition of designing algorithms that produce human-like sketches of scenes --- whether from 3D models (see \citet{benard2019line} for a survey) or directly from images \citep{chan2022learning, vinker2022clipasso}.
These methods often seek to capture human intuitions about which contours are the most important ones to depict: for example, when depicting geometry \citep{cole2008people, cole2009well}, maps \citep{agrawala2001rendering}, and abstract concepts \citep{vinker2022clipasso}.

Insights about these intuitions are doubly valuable: they can be applied not only to \emph{generate} human-like sketches, but also to \emph{understand} them in a human-like way. Much work in graphics and HCI has applied such insights to design intuitive sketch-based interfaces for 3D modeling \citep{igarashi2006teddy, karpenko2006smoothsketch, davis2006sketching, tversky2003sketches, olsen2009sketch}. Studying human sketching remains an active research area in cognitive science \citep{drawing-as-tool}, AI \citep{mukherjee2024seva}, and graphics \citep{hertzmann2024new}, and cognitive insights about human sketching continue to influence work in the graphics community \citep{sarukkai2024block, chandra2024coggraph}.

In this paper, we seek to expand the scope of this research to the domain of \emph{sound}. Until now, the graphics community has largely studied \emph{realistic} sound synthesis, either by physical simulation \citep{dobashi2003real, zheng2011toward, raghuvanshi2010precomputed, james2006precomputed, o2002synthesizing, van2001foleyautomatic, liu2019physically, xue2023improved, wang2018toward} or by data-driven methods \citep{owens2016visually, cardle2003sound, kong2020diffwave}. To the best of our knowledge, our work is the first to study non-phonorealistic sound synthesis.

We study human intuitions about vocal imitation not only to generate human-like vocal imitations, but also to understand vocal imitations well enough to build new intuitive ``sketch-based'' interfaces for sound artists.
Such interfaces were previously envisioned by \citet{ekman2010using} and \citet{lemaitre2014effectiveness}. Prior work in human-computer interaction has laid the foundations for such interfaces by measuring human vocal imitation abilities \citep{cartwright2015vocalsketch, lemaitre2016vocal}, and recent work on ``onomatopoeia-based synthesis'' has sought to build such interfaces using deep learning on paired sound/onomatopoeia data \citep{ohnaka2023visual, takizawa2023synthesis, okamoto2022onoma}. Our work instead approaches this problem from first principles of vocal imitation, and thus does not require a paired dataset of sounds and vocal imitations.


To design our vocal imitation system, we draw on the theory of non-photorealistic depiction in graphics: specifically,
\citet{durand2002invitation}'s framework, which conceives of depiction as optimizing a stimulus to best induce a desired percept in the audience's mind.
Recent work in graphics has applied this framework to a variety of domains: for example, to optimize sketches that best convey depth \citep{chan2022learning}, to optimize visual illusions that evoke interesting percepts \citep{chandra2022designing}, and to optimize animations that best convey physical and social attributes of characters \citep{chandra2023acting, chen2024intervening}.
We take the same approach here: we model vocal imitation as optimizing a vocal utterance to best communicate the desired sound to (a model of) the listener. We discuss our cognitively-inspired model of the listener, and how we optimize over it, in Section~\ref{sec:rsa}.

\section{Method}

We focus on vocal imitations of sound \emph{textures} --- sounds whose perceptual features remain stable over long time scales \citep{saint1995analysis}.
Our goal is to design a method for doing two tasks:
\begin{enumerate}
    \item Speaking: given a short audio clip of a sound texture, which we will call the \textbf{referent}, produce a human-like vocal imitation of that sound texture, which we will call an \textbf{utterance}.
    \item Listening: given a real human vocal imitation of a sound texture (an \textbf{utterance}), infer which sound texture the speaker was imitating (the \textbf{referent}).
\end{enumerate}
We will say that utterances, which we will denote with $u$, lie in $U$, the space of possible human vocalizations. Referents, which we will denote with $r$, lie in $R$, the space of all sound textures. Importantly, $U \subsetneq R$: possible human vocalizations are only a very sparse subset of all sound textures (much like how line drawings are only a very sparse subset of all possible images). This is what makes vocal imitation so challenging and so interesting.

We will operationalize our two tasks as two probability distributions to be approximated:
\begin{enumerate}
    \item $p_S(u \mid r)$ is the distribution over utterances that a human speaker ($S$) might make to communicate a given referent.
    \item $p_L(r \mid u)$ is the distribution over referents that a human listener ($L$) might infer upon hearing a given utterance.
\end{enumerate}
We will start in Section~\ref{sec:naive} by describing a simple model of the speaker $p_S(u \mid r)$ that chooses $u$ based on how well $u$ matches the perceptually-relevant auditory features of $r$. However, we will see that this on its own does not match human intuitions well. In Section~\ref{sec:rsa}, we will explain why, and solve the problem by having the speaker take into account the listener's intuitions. This gives a natural method of jointly computing the speaker's $p_S(u \mid r)$ and the listener's $p_L(r \mid u)$ side-by-side.

Source code for the models described in this paper is available at
\url{https://github.com/matthewcaren/vocal-imitation}.

\subsection{Warm-up: vocal imitation by direct optimization}\label{sec:naive}

\begin{figure*}
    \centering
    \includegraphics[width=0.8\linewidth]{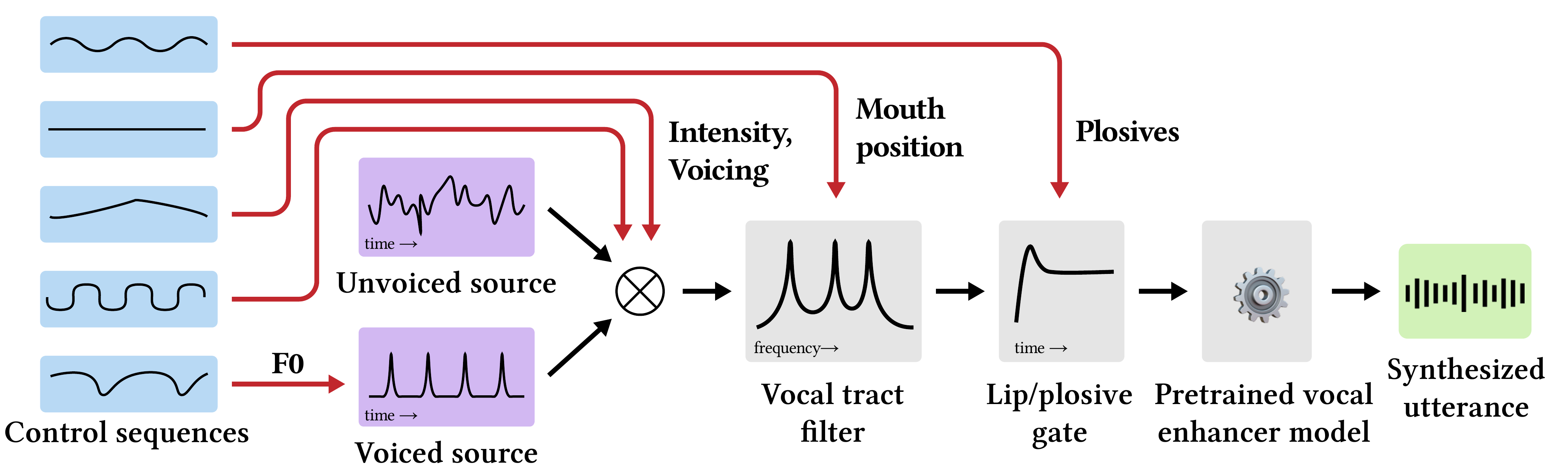}
    \caption{Schematic diagram of our source-filter model of the vocal tract (see Section~\ref{sec:naive}).}
    \label{fig:vocal-tract}
\end{figure*}

We will start by producing vocal imitations by optimizing utterances to match the referent's perceptually-relevant auditory features --- a strategy that has traditionally worked well for both auditory and visual texture synthesis \citep{mcdermott-texture-2009, mcdermott-texture-2011, heeger1995pyramid, gatys2015texture}.

\subsubsection{Defining the space of utterances}

Our first task is to define $U$, the space of possible utterances. To do this, we create a controllable model of the human vocal tract, which takes as input a control sequence (representing lip/tongue movements, etc.) and synthesizes a corresponding human-like vocalization.
Before reading on, we encourage readers to play with on-line demos of controllable vocal tract models, such as the ``Pink Trombone'' \citep{pinktrombone}, to gain intuition for how the human vocal tract responds to various controls.

We use a standard source-filter model (see Figure~\ref{fig:vocal-tract}), which simulates the vocal tract as a \textit{source} signal that models the oscillation of the vocal folds and aspiration/frication, along with a \textit{filter} that models the resonances of the oral and nasal cavities \cite{sfmodel-fant}. The source is modeled as an impulse train (for voiced sounds) as well as broadband noise (for unvoiced sounds). The filter is modeled as a bank of resonant LTI filters tuned to match resonant formant frequencies created by various tongue/mouth positions, as well as a bank of envelope generators that model plosive consonants.

We chose the source-filter paradigm because it aligns closely with the physical behavior of the vocal tract \cite{sf-source, sf-fric, sf-phonetics-tb}. Alternative approaches include concatenative synthesis \cite{concat-voice-synth} (as used by products like Siri and Cortana) and deep learning-based systems \citep{wavenet}. We did not use these approaches because they are typically constrained to a specific narrow domain, e.g.\ speech in a particular language. The source-filter approach allows us to flexibly synthesize arbitrary phoneme sequences that span much more of the space $U$. It is also computationally efficient and has interpretable control parameters. However, it is not without limitations (see Sec~\ref{sec:limitations}).

We implemented our vocal tract model in the \textsc{Faust} language \cite{sf-faust} based on the \textsc{Faust-STK} modal model \cite{sf-fauststk}, and tuned it to generate utterances for stereotypically masculine and stereotypically feminine voices.
Finally, we used a pretrained off-the-shelf vocal enhancer model \citep{resemble} to fix-up the generated sounds to sound less robotic and bridge the ``uncanny valley.'' (The use of this vocal enhancer model is purely cosmetic. To give readers a qualitative sense of the effect of this model, we include some ``before/after'' examples in the supplementary materials.)

To define $U$, we generated a large set of utterances from our vocal tract model by discretizing the space of control sequences given to our model. For each of 5 parameters (fundamental frequency $F_0$, overall loudness, vowel type, plosive gate, and voicedness), we modulated it according to one of 11 patterns (constant, sinusoidal/sawtooth with various frequencies, and random walk). This led to an utterance space of $|U|=11^5=161,051$ utterances.

\subsubsection{Comparing $u$ and $r$ based on auditory features} Now that we have defined the utterance space $U$, the next step is to search for a $u$ that ``sounds like'' a given $r$. How can we compute how much an utterance $u$ sounds like a referent $r$? To model this, we extract and compare a variety of commonly-used perceptually-relevant auditory features $\phi_i$ of $u$ and $r$: the mean and standard deviation of the sound's spectral flatness, spectral centroid, spectral peak, and RMS loudness, as well as their derivatives \citep{mullermusicprocessing}. (Our method is robust to the specific choice of features --- see Section~\ref{sec:res-ablations}.)

Using $\phi$, we can finally make a first attempt at vocal imitation by feature matching. Given a referent $r$, we select an utterance $u$ by performing a softmax over the cosine similarity between the features of $r$ and $u$:
\begin{equation}\label{eqn:naive-S}
p_S(u \mid r) \propto \exp\left( \beta \cdot \frac{\phi(r) \cdot \phi(u)}{|\phi(r)||\phi(u)|} \right).
\end{equation}
(Throughout this paper, we will define probabilities in terms of proportionalities as a notational convenience to avoid writing the normalizing constant that makes the probabilities sum to 1. In general, the statement $p(x) \propto f(x)$ is the same as $p(x) = f(x) / \Sigma_x f(x)$.)

We can model the listener exactly the same way: given an utterance $u$, the listener infers referents $r$ by performing a softmax over the cosine similarity between features of $r$ and $u$:
\(
p_L(r \mid u) \propto \exp\left( \beta \cdot \frac{\phi(r) \cdot \phi(u)}{|\phi(r)||\phi(u)|} \right)
\).
We have the listener select among referents in the FSD50K dataset \cite{fsd50k}, which contains 51,197 audio clips from Freesound database. That is, $|R|=51,197$.

\subsubsection{Limitations of feature matching}\label{sec:naive-issue}
For some sounds, such as the rustling of leaves, $p_S$ predicts utterances that indeed sound quite reasonable. However, for many other sounds, the results do not match our intuitions very well. For example, consider the sound of a speeding motorboat, which is characterized by loud broadband noise (from the splashing of the water) as well as a low-pitched rumble (from the engine). Because the broadband noise is the most perceptually salient feature, our system produces a loud unvoiced fricative (``shhhhh''). However, when people are played an audio clip of a speeding motorboat and asked to vocally imitate the sound, they intuitively ignore the loud broadband noise and instead choose to imitate the low-pitched rumble (``woh-woh-woh-woh'').

\begin{figure}
    \centering
    \includegraphics[trim={0 8.9cm 6.5cm 0},clip,width=\linewidth]{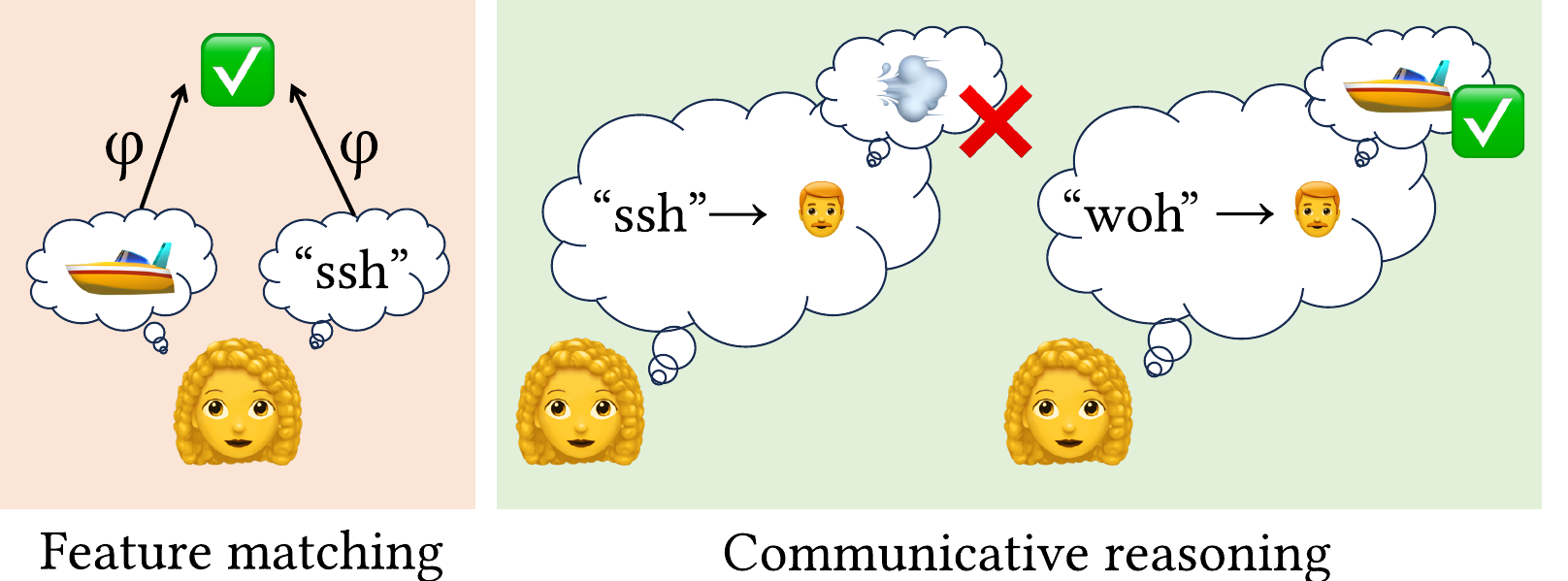}
    \caption{The problem with the feature matching baseline (Section~\ref{sec:naive-issue}). The sound of a motorboat is predominantly the water's loud broadband noise, well-matched by ``ssh'' (left). However, a speaker trying to imitate a motorboat is likelier to imitate the engine's rumble (``woh'') because it would be more distinctive to a listener --- ``shh'' could be mistaken for wind.}
    \label{fig:feature-vs-comm}
\end{figure}

This illustrates a key point about human vocal imitation (Figure~\ref{fig:feature-vs-comm}): we do not choose to imitate the most perceptually salient features of the referent. Rather, we choose to imitate its most \emph{informative} features, in order to help the listener best distinguish the referent from alternate possibilities. The broadband noise is a poor choice because many phenomena in the world produce loud broadband sounds (e.g.\ leaves, wind, ...), and so a listener would have high uncertainty over the referent the speaker intended. On the other hand, the low-pitched rumble is a very distinctive aspect of the sound of a speeding motorboat, and narrows down the space of alternate hypotheses considerably for the listener.
In the next section, we will augment our system to perform such communicative reasoning.

\subsection{Adding communicative reasoning to our system}\label{sec:rsa}

The phenomenon we describe above is a classic example of \emph{pragmatic reasoning} in human communication: a subject that has been well-studied by cognitive scientists and linguists over the past decades \citep{grice1975logic}.
One particularly successful computational account of human pragmatic reasoning was given by \citet{goodman-RSA-2012}, who considered recursively modeling speakers as reasoning with respect to imagined models of listeners (who might themselves be imagining models of speakers). Remarkably, \citet{goodman-RSA-2016} found that describing communication scenarios in this way predicts human communicative behavior very well. They call their framework the ``Rational Speech Acts'' framework or ``RSA'' (not to be confused with the cryptographic algorithm).

RSA's success in modeling human communication has motivated researchers across disciplines to develop cognitively-grounded intuitive interfaces for human-computer communication \citep{acquaviva2022communicating, vaithilingam2023usability, pu2020program}. Here, we will apply RSA to improve our vocal imitation model.

\subsubsection{Applying RSA to vocal imitation}

In terms of a recursive RSA model, our feature-matching system is a ``base-case'' speaker who is ambivalent to the listener. In reality, however, human speakers reason recursively about how the listener might interpret their speech, in order to cooperatively provide the most communicative utterance. To incorporate this insight, we consider a ``level-2'' speaker $S_2$ who strategically chooses an utterance $u$, not based on matching features, but rather based on the probability that a base-case listener would choose the correct referent if they heard $u$: \begin{equation}\label{eqn:rsa-nocost-S}
p_{S_2}(u \mid r) \propto \exp(\beta \cdot p_L(r \mid u)).
\end{equation}
In the same spirit, we can now define a level-2 \emph{listener} $L_2$ who infers referents using Bayes' rule, assuming that the utterance came from a level-2 speaker:
\(
    p_{L_2}(r \mid u) \propto p_{S_2}(u \mid r) \cdot p(r)
\). (Here $p(r)$ is an arbitrary Bayesian prior over referents, typically set to be uniform.)

In principle, we can continue iterating this process to obtain a sequence of higher-level $S_k$ speakers and $L_k$ listeners \emph{ad infinitum}. However, in practice, for small $k$ (even just 2 or 3) these computations quickly converge to probability distributions that match human intuitions extremely well \citep{goodman-RSA-2012, goodman-RSA-2016}. Hence, we stop at $S_2$ in our system.

\subsection{Incorporating the speaker's costs and utilities}\label{sec:rsa-costs}

In addition to recursive reasoning, the RSA framework also provides a principled way for us to take into account another key aspect of human vocal imitation: the \emph{costs} and \emph{utilities} associated with producing a given utterance.

\begin{figure}
    \centering
    \includegraphics[trim={0 8.9cm 6.5cm 0},clip,width=\linewidth]{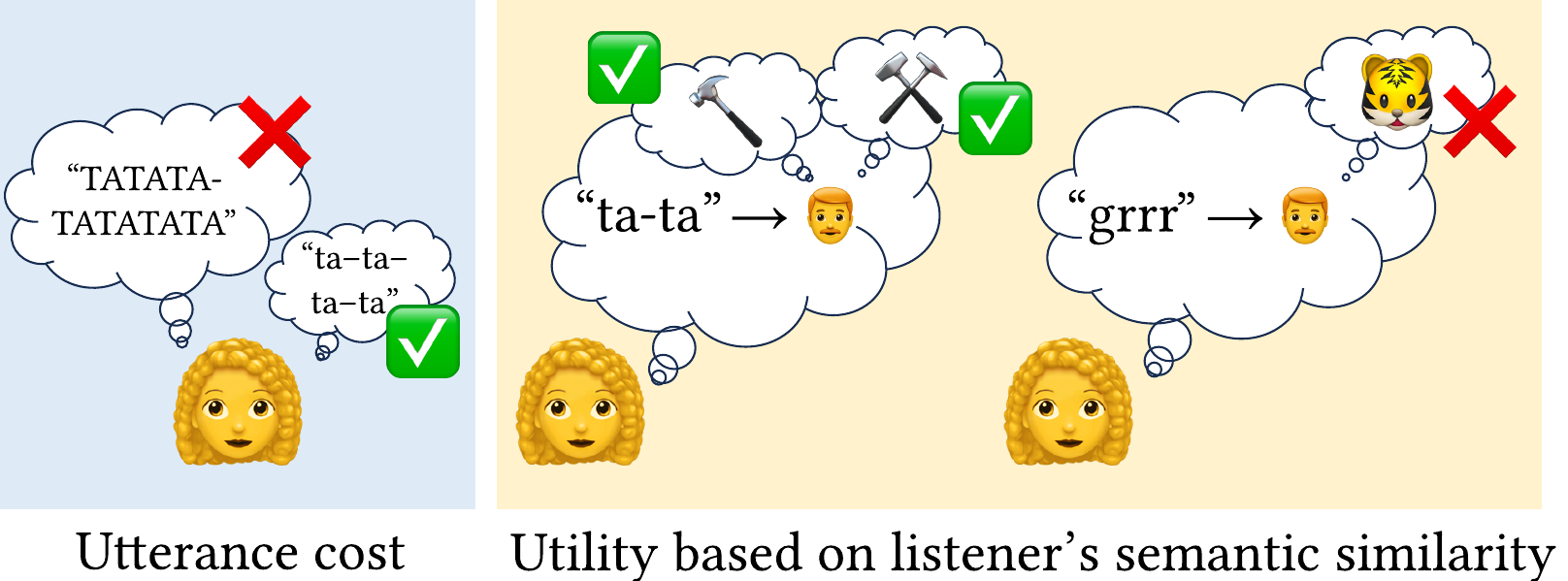}
    \caption{Incorporating costs and utilities (Section~\ref{sec:rsa-costs}). A speaker imitating a jackhammer might go with a softer, slower sound that is easier to make, and might be okay if the listener infers a different tool (but not, say, a tiger).}
    \label{fig:cost-utility}
\end{figure}

The speaker's \emph{cost} of producing a given utterance could come from many sources. For example, imagine vocally imitating a jackhammer's fast repetitive sound as \emph{ta-ta-ta-ta}. It might be possible to perfectly reproduce that tempo by moving your tongue very rapidly, but it would be challenging and require a lot of focus. Making a slightly \emph{slower} utterance might still adequately communicate ``jackhammer'' to the listener. In this case, speakers should rationally opt for the slower utterance (Fig~\ref{fig:cost-utility}, left).

Similarly, the speaker's \emph{utility} of an utterance could depend on the speaker's goals. For example, when vocally imitating a jackhammer to emphasize the loud construction noise that woke them up that morning, a speaker would be most satisfied if the listener infers a jackhammer. But they might still be partially satisfied if the listener infers a different power tool, like a drill. On the other hand, they would be much less satisfied if the listener infers a completely different category of sound, such as the growl of a tiger (Fig~\ref{fig:cost-utility}, right).

To model such costs and utilities, we generalize our model of the speaker from Equation~\ref{eqn:rsa-nocost-S} to
\begin{equation}\label{eqn:rsa-full-S}
p_{S_2}(u \mid r) \propto \exp(\beta \cdot (V_2(u, r) - c(u))),
\end{equation}
where $c(u)$ is the cost of producing utterance $u$, and $V_2(u, r)$ is the expected utility to an $S_2$ speaker of producing utterance $u$ to communicate referent $r$.
We model the cost $c(u)$ of an utterance as proportional to (a)~the rate at which the control parameters of the vocal tract model are varied, and (b)~the proportion of the utterance spent at the ``extremes'' of vocal tract model parameters. This assigns high cost to utterances that are very rapid, very loud, and very high/low-pitched.

We model the speaker's utility $V_2(u, r)$ of producing utterance $u$ for referent $r$ in terms of how semantically-similar the listener's \emph{inferred} referent $r^\prime$ is to $r$. To measure this semantic similarity, we use the fact that audio clips in the FSD50K dataset are classified according to the AudioSet ontology \cite{audioset}, a rich hierarchical classification scheme for sounds. For example, AudioSet might classify a particular sound of a cat under \textit{Animal~$\gg$ Pets~$\gg$ Cat~$\gg$ Meow~$\gg$ Meow~\#4}. The full ontology has 144 leaf nodes and 56 intermediate nodes.
Let $\Delta(r_1, r_2)$ be the number of matching levels between the classification of $r_1$ and $r_2$. For example, \textit{Animals~$\gg$ Pets~$\gg$ Dog~$\gg$ Bark~$\gg$ Bark~\#7} has two matching levels to \textit{Meow~\#4} (\textit{Animals~$\gg$~Pets}). We model the speaker's utility as the expected $\Delta$ between the referent $r$ and the listener's inferred $r^\prime$, where the expectation is taken over $r^\prime \sim p_{L}(r^\prime \mid u)$:
\[
V_k(u, r) = \mathop{\mathbf{E}}_{r^\prime}  \left[~\Delta(r, r^\prime)~\right]
= \sum_{r^\prime \in R}  \Delta(r, r^\prime) \cdot p_{L}(r^\prime \mid u).
\]
Notice the recursive call to the baseline $L$ listener, which represents the speaker reasoning about the listener.

\subsection{Interim summary}\label{sec:interim-summary}

In this section, we built up a model of human vocal imitation step by step from first principles of communication, in three stages:
\begin{enumerate}
    \item We started by directly matching perceptually-relevant auditory features of utterances produced by a model of the vocal tract (Eqn~\ref{eqn:naive-S}, Sec~\ref{sec:naive}). We call this the \textbf{baseline} model.
    \item We then applied a cognitive theory to model the speaker and listener' communicative reasoning (Eqn~\ref{eqn:rsa-nocost-S}, Sec~\ref{sec:rsa}). We call this the \textbf{communicative-only} model.
    \item Finally, we added terms to model the speaker's costs and utilities (Eqn~\ref{eqn:rsa-full-S}, Sec~\ref{sec:rsa-costs}). We call this our \textbf{full} model.
\end{enumerate}
These models are all relatively lightweight: we ran our computations on a 2021 M1 MacBook Pro, and the full method (the most computationally expensive) took only 14 minutes to predict vocal imitations for all 51,197 audio clips in FSD50K.

In the next section, we will evaluate the vocal imitations produced by these models.

\section{Evaluation}

Recall that our goal was to design a system that can produce and understand human-like vocal imitations. In this section, we evaluate our method on four criteria: how human-like its utterances are (Sec~\ref{sec:res-humans}), how human judges rate its utterances (Sec~\ref{sec:res-users}), how well it can flexibly adapt to new constraints (Sec~\ref{sec:res-whisper}), and how well it can understand imitations (Sec~\ref{sec:res-retrieval}). We compare our \textbf{full} method against the two ablations from Sec~\ref{sec:interim-summary}: \textbf{baseline} and \textbf{communicative-only}. All of the vocal imitations discussed in this section are available in the supplementary materials, and online at \url{https://matthewcaren.github.io/vocal-imitation/}. 

\subsection{Our method produces human-like vocal imitations}\label{sec:res-humans}

\begin{figure*}[t]
    \centering
    \includegraphics[trim={0 0.3cm 0cm 0},clip,width=\linewidth]{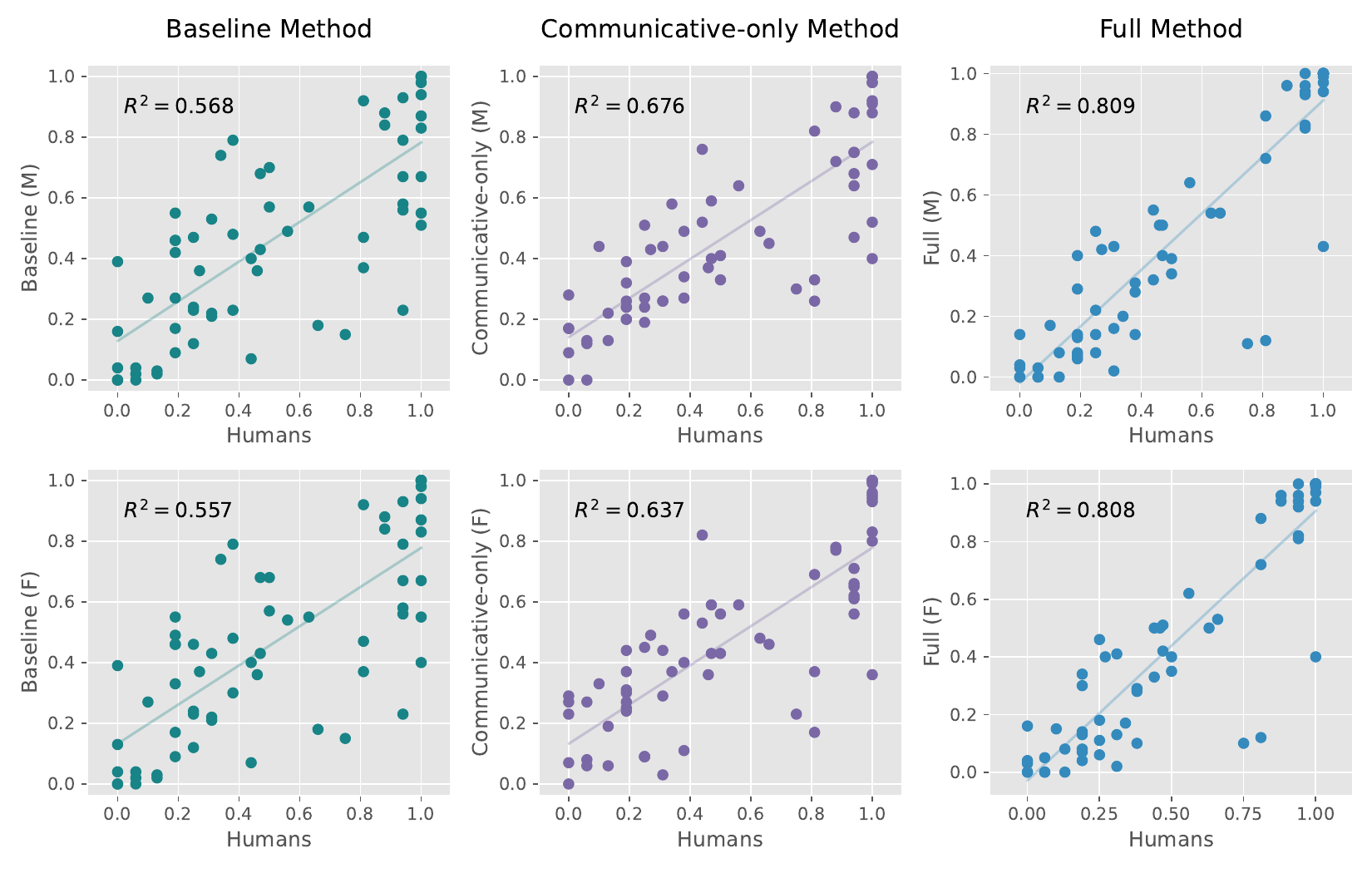}%
    \caption{Vocal imitations generated with communicative reasoning are significantly more human-like than those produced by the feature-matching baseline. Averaged across stereotypically male and female voices, the correlation of phonetic features between \textbf{communicative-only} and human utterances is 0.65, as compared to 0.56 for the \textbf{baseline} model. Modeling the speaker's costs and utilities, as in our \textbf{full} method, further increases the correlation to 0.81.}
    \label{fig:scatter}
\end{figure*}

To compare our method to humans, we used the VocalImitationSet dataset of experimentally-collected human-generated vocal imitations of sounds in the AudioSet ontology \citep{vocalimitationset}. We started by randomly sampling $N=16$ referents from $R$. For each referent, we randomly sampled $k=16$ human-generated vocal imitation utterances from VocalImitationSet. Separately, we ran our three methods on each of the $N=16$ referents to get each method's probability distributions over utterances for each referent.

Next, we coded each utterance for four key phonetic features: whether it was \emph{voiced}, whether it contained \emph{stops}, whether the vowels were \emph{open} (i.e. low), and whether they were \emph{fronted}.
To be clear, these \emph{phonetic} features are different from the \emph{auditory} features used to measure perceptual similarity within our system. The phonetic features represent the strategies people use to vocally imitate, not the actual sounds they produce. For example, we consider it a success if both humans and our method produce an ``oooh'' sound for a given referent, even if the two ``oooh'' sounds vary in absolute pitch. On the other hand, we do \emph{not} consider it a success if both methods produce sounds of exactly the same pitch, but humans produce ``oooh'' while our method produces ``aaah'' (more phonetically \emph{open}).

For the $N\times k = 256$ human-generated utterances, we manually coded the 4 phonetic features. For our system's synthetic utterances, we coded these features automatically, because we knew the control sequences used to synthesize each utterance. Finally, we compared the distribution of features in human-generated utterances to the distribution of features predicted by our methods.

\begin{figure*}
    \includegraphics[width=0.32\linewidth]{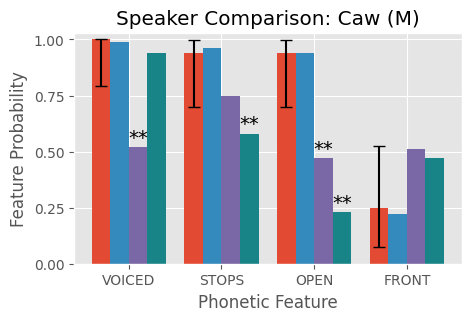}~~~
    \includegraphics[width=0.32\linewidth]{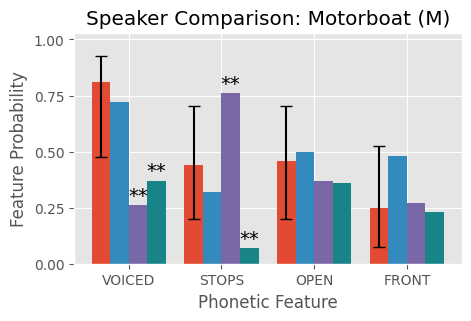}~~~
    \includegraphics[width=0.32\linewidth]{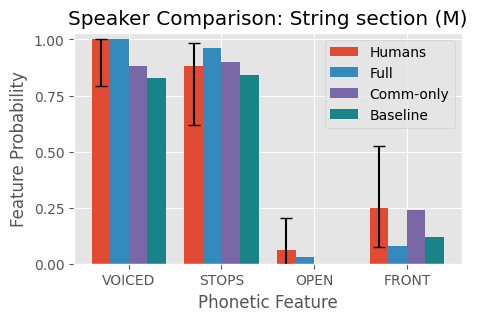}%
    
    \includegraphics[width=0.32\linewidth]{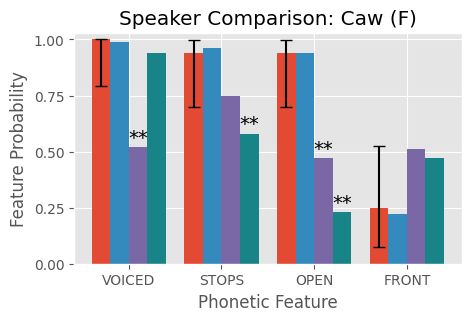}~~~
    \includegraphics[width=0.32\linewidth]{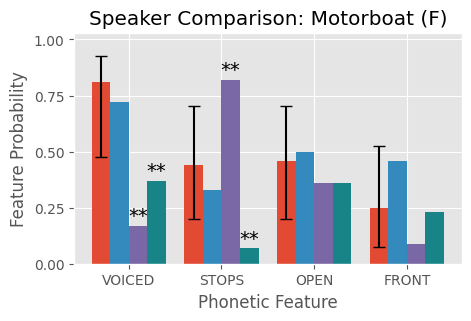}~~~
    \includegraphics[width=0.32\linewidth]{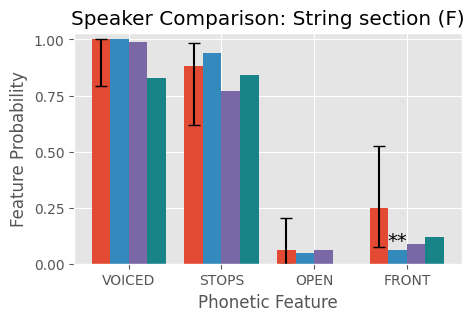}%
    \caption{Our full method successfully and robustly predicts high-level phonetic features of human-generated vocal imitations. In comparison, the feature-matching baseline, and the communicative-only ablation, differ significantly from humans in many cases. The top row shows results with our vocal tract model tuned for stereotypically masculine voices, and the bottom row shows results for stereotypically feminine voices. Error bars are 95\% confidence intervals, and stars denote statistically significant deviations from humans.}
    \label{fig:bars}
\end{figure*}

Figure~\ref{fig:bars} shows some examples of feature frequencies for individual referents. Our full method generally matches humans very closely. However, without communicative reasoning, the results begin to diverge from humans. For example, when imitating a crow's ``caw,'' humans and our full method almost always produce an \emph{open} vowel (IPA:~[a]), while the baseline feature-matching method typically produces a \emph{closed} vowel (IPA:~[i]). This is because the [i] sound's high-frequency components closely match the referent sound sample's auditory features. The communicative-only method places somewhat higher weight on [a], but only our full method is able to reason in a human-like way and confidently choose [a].

Figure~\ref{fig:scatter} shows this data aggregated across all $16$ referents that we considered for this evaluation: each point corresponds to a bar as in Figure~\ref{fig:bars}, for a total of $16\times4 = 64$ points per scatterplot. The baseline method has relatively low correlation with humans ($r^2=0.57, 0.56$ respectively for stereotypically masculine and feminine voices), and the communicative-only method is slightly higher ($r^2=0.68, 0.64$ respectively). Our full method, however, matches human behavior strikingly well ($r^2=0.81$ for both). This is particularly remarkable because we never fit any parameters of our model to actual human-generated vocal imitations.

When we inspected the outliers in the scatterplots, we found that they corresponded to cases where people reached for culturally-evolved linguistic conventions (e.g. onomatopoeia) rather than direct vocal imitations. For example, one of the outliers is for the referent ``heartbeat'' and the phonetic feature of voicedness. People often imitate heartbeats with the voiced words ``ba-boom, ba-boom,'' even though the referent is actually short and broadband, and thus better matched by unvoiced plosives. Our method diverges from humans by producing unvoiced plosive (``thup-thup''). But this is due to an arbitrary cultural artifact --- indeed, in some eastern European languages like Latvian and Russian, the word for a heartbeat's sound \emph{does} begin with a ``th'' sound (IPA: [$\theta$]).

\subsection{Humans prefer our full method's utterances to the baseline and communicative-only utterances}\label{sec:res-users}

\begin{figure}
    \centering
    \includegraphics[trim={0 0.4cm 0cm 0},clip,width=\linewidth]{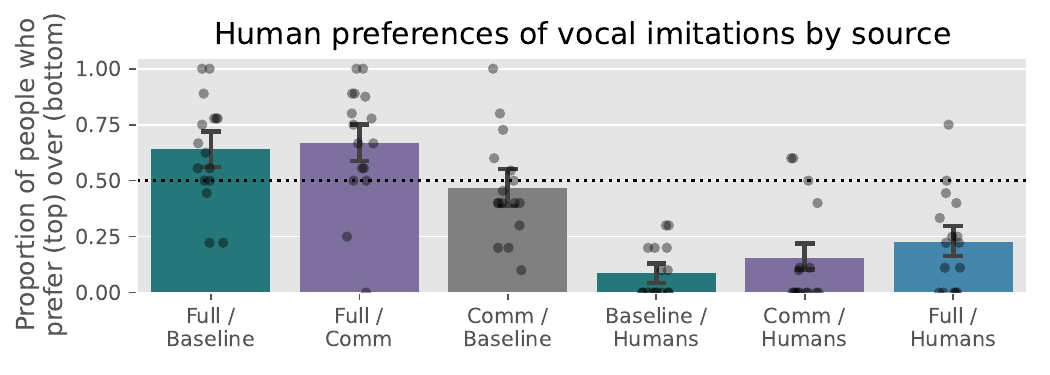}
    \caption{Human preferences of vocal imitations (Sec~\ref{sec:res-users}). The dotted line indicates chance (no preference). Vocal imitations generated by our full method are preferred over the feature-matching baseline (bar~1) and communicative-only method (bar~2). Notably, 25\% of the time people even prefer our full method's utterances over actual \emph{human-generated} imitations (bar~6). Error bars are 95\% confidence intervals.}
    \label{fig:user-study}
\end{figure}

Next, we studied how humans respond to the vocal imitations produced by our method.
We recruited $N=60$ participants for an IRB-approved on-line user study on the Prolific platform (Figure~\ref{fig:exp1}). Participants were first played an audio clip of a referent. Then, they were played two vocal imitations from different sources, which were chosen at random among \textbf{baseline}, \textbf{communicative-only}, \textbf{full}, and a randomly-sampled \textbf{human} vocal imitation from VocalImitationSet, for a total of $\binom{4}{2}=6$ possible pairs of sources (for synthesized imitations, we selected the top-1 predicted utterance). Participants selected which of the two vocal imitations they thought better matched the referent. Participants completed 16 rounds, one for each referent considered in the previous section, in randomly-shuffled order. They were paid \$2.50 for an average rate of \$15/hour.

Our results are shown in Figure~\ref{fig:user-study}. Each of the 6 bars represents preference between one pair of sources, and dots show average preference grouped by referent (hence, there are 16 dots on each bar).
There are two key takeaways from these results. First, people consistently prefer our full method's utterances over the baseline and the communicative-only utterances (bars 1--2, above chance), though they do not have a significant preference between the baseline and communicative-only utterances (bar 3, at chance).
Second, even when compared against \emph{human-generated} utterances, people prefer our full method's utterances around 25\% of the time (bar 6). Looking at the individual referents (dots), we can see that our full method is competitive with humans (near or beyond chance levels) for several individual referents. Again, we emphasize that this is particularly remarkable for a system that was never trained on human vocal imitations --- our full method's ability to persuade people derives entirely from the design of our model. Indeed, communicative-only utterances fare worse against human-generated utterances, and the baseline utterances are the least persuasive (bars 4--6, upward trend).

\subsection{Our method is flexible to novel constraints}\label{sec:res-whisper}
Not only can people use their vocal tract to imitate sounds, they can also do so under a variety of additional constraints. For example, in a quiet library, people might produce a \emph{whispered} vocal imitation---a vocal imitation constrained to unvoiced sounds.
Our method can easily be adapted to such constraints. Recall the speaker's cost term $c(u)$ --- we can simply increase the cost of voiced utterances to $+\infty$, which has the effect of removing voiced utterances from consideration. We provide some examples of whispered vocal imitations in the supplementary materials. Notably, the feature-matching baseline struggles in this regime because of the extreme domain shift between referents and whispered utterances.

\subsection{Our method can be used to understand vocal imitations}\label{sec:res-retrieval}

Just like how the computer graphics and computer vision communities have designed several systems for retrieving images from sketches \citep{sti-bagoffeatures, sti-gfhog, sti-sketch2photo, sti-sketchygan, sti-thatshoe, sti-color, sti-3d}, a smaller community has designed systems for retrieving sounds from vocal imitations \citep{vi-querying-freesound, vi-autoencoder, blancas2014sound}. These existing systems rely primarily on feature matching, analogous to our baseline method.

As we discussed in Section~\ref{sec:rsa}, our method naturally provides a way to model both speakers ($p_S$) and listeners ($p_L$). We evaluated our three methods' corresponding listeners on the task of retrieving \emph{animal sounds} from vocal imitations in VocalImitationSet. For each of 30 animal sounds in the AudioSet ontology (e.g.\ cat purr, dog growl, frog croak), we gathered 16 human-generated vocal imitations from VocalImitationSet to build a test set of size $30\times16 = 480$. We tested our three methods on this benchmark, and computed each method's top-1 accuracy for the full 30-way classification task, as well as a coarser 3-way classification task where the options were aggregated into ``domestic animals,'' ``livestock,'' and ``wild animals.''

\begin{figure*}
    \centering
\includegraphics[width=0.45\linewidth]{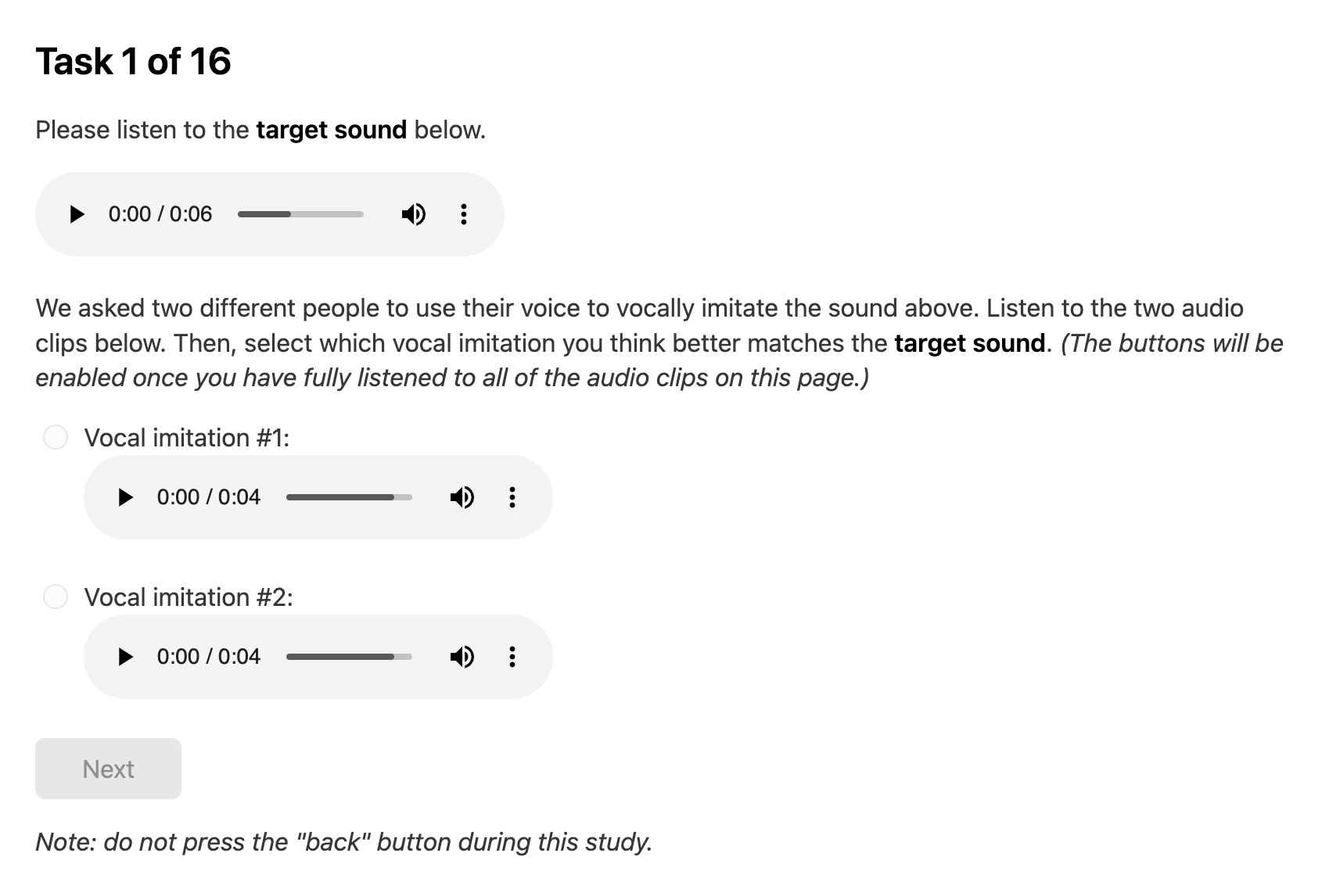}\hfill
\includegraphics[width=0.45\linewidth]{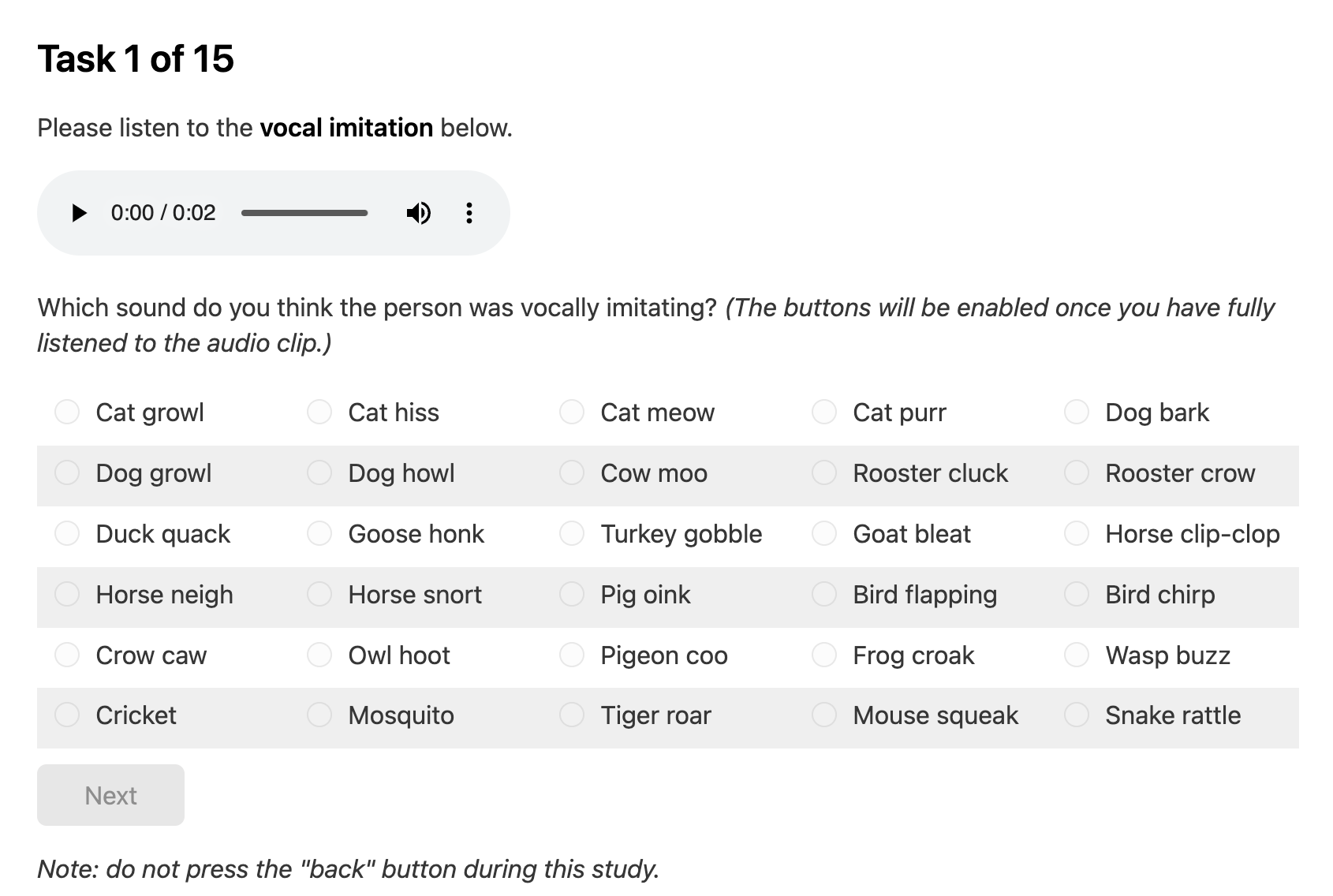}
    \caption{Screenshots of the interfaces used in the experiments described in Sections~\ref{sec:res-users} and~\ref{sec:res-retrieval}.}
    \label{fig:exp1}
\end{figure*}

We compared our methods to humans by running another IRB-approved human subject study to measure their performance on this task (Figure~\ref{fig:exp1}). We recruited 32 participants on Prolific. Each participant first confirmed that they knew of all 30 animal sounds. For sounds they did not know, they listened to a sample audio clip. Then, they listened to 15 human-generated vocal imitations, and for each one guessed (among 30 choices) what animal sound they thought the speaker was imitating. Participants were paid \$2.50 for an average rate of \$15/hour. We collected $32\times 15=480$ guesses, one for each of the $30\times 16=480$ vocal imitations in our test set.

For the 3-way classification task, our full method achieves a top-1 accuracy of 58\%, while humans achieve 70\%. In comparison, the baseline achieves 46\%, and the communicative-only method achieves 55\%.
For the full 30-way classification task, our full method achieves a top-1 accuracy of 18\%, and a top-5 accuracy of 53\%. This is a challenging task: humans only achieve a top-1 accuracy of 42\%. Like the full method, the communicative-only method also achieves a top-1 accuracy of 18\%, while the baseline achieves 7\% (top-5: 31\%).

\begin{figure*}
    \centering
    \includegraphics[width=\linewidth]{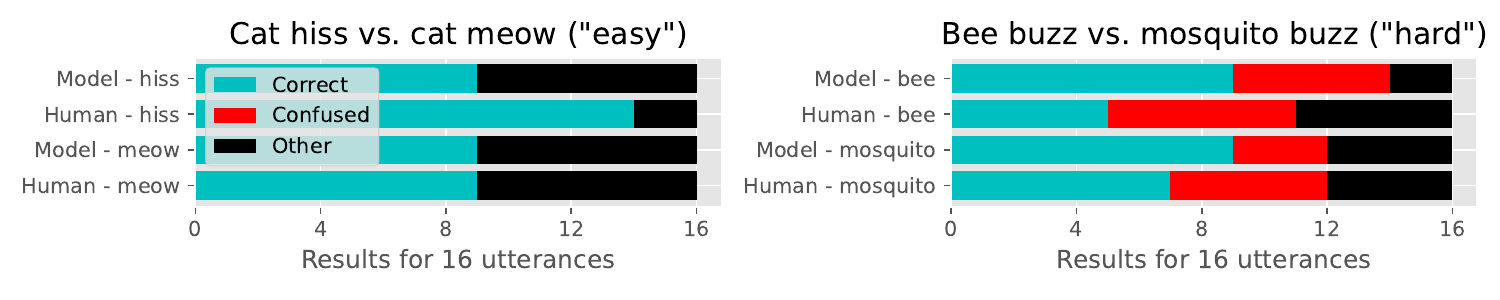}
    \caption{Our full method's listener model makes similar patterns of errors as humans do. For example, neither our model nor humans ever confuse a cat's hisses for meows (left). However, they both frequently confuse a bee's buzz with a mosquito's (right). See Section~\ref{sec:res-retrieval} for a discussion.}
    \label{fig:confusion}
\end{figure*}

Interestingly, our full method and humans make similar patterns of errors on this task. For example, like humans, our method confuses imitations of ``bee buzz'' and ``mosquito buzz,'' but does not confuse imitations of ``cat hiss'' and ``cat meow'' (see Figure~\ref{fig:confusion}). This is preliminary evidence, and more work is needed to precisely characterize the kinds of errors that humans and our model make.

\subsection{Our method is robust to the choice of auditory features}\label{sec:res-ablations}

Finally, we evaluated our full method's sensitivity to the specific choice of low-level auditory features selected in Section~\ref{sec:naive}, by lesioning 6 of the 19 features (2 each at random from loudness, pitch, and timbre). The resulting $S_2$ speaker in the full method has a correlation with humans of $r^2 = 0.74$, down only slightly from $0.81$ with all 19 features. This shows that communicative reasoning is robust to the specific choice of base features, and reflects a general feature of human communication known as \emph{pragmatic strengthening} \citep{goodman2015probabilistic}.

\section{Limitations and Future Work}

\paragraph{Where does our method diverge from humans?} \label{sec:limitations}
There are three main situations where our method diverges from humans.
First, as discussed in Section~\ref{sec:res-humans}, our method cannot predict culturally-evolved linguistic conventions (e.g.\ ``ba-boom'' for heartbeat).

Second, because we focus on vocal imitations of sound textures, our method also cannot handle sounds that have higher-order temporal structure, like speech and music. Extending our method beyond sound textures could enable a variety of new applications, such as ``query-by-humming'' (e.g.\ Soundhound), which is currently typically implemented by feature-matching \citep{qbh-cornell, qbh-morerepresentations, qbh-google, qbh-survey, qbh-crowdsource}.

Finally, our vocal tract model cannot yet produce certain consonants, such as [z], which leads it to diverge from humans in some cases. For example, our method produces a long whining sound, rather than a buzzing sound, to imitate a bee.
More generally, our vocal tract model has much scope for improvement.
While there is a long history of research on physically-based controllable modeling of human voice \citep{vocoder-history}, we have yet to achieve a general-purpose model \citep{vocaltractlab} --- and while recent realistic neural voice models \citep{voicebox, virtuoso} excel at \emph{speech} modeling, they cannot produce the full range of sounds that humans can make. 
A related limitation of our method is its reliance on a discretized utterance domain. Extending our method to a continuous utterance domain, for example as proposed by \citet{andreas-klein}, would undoubtedly improve results. However, we generally found that our discretization was fine enough to cover our vocal tract model's domain with sufficient granularity to qualitatively match most human-generated utterances of sound textures.

\paragraph{Future work}
Besides extending our methods in the ways discussed above, there are two broader directions we are pursuing.
First, we are interested in methods for other forms of non-phonorealistic sound synthesis. For example, electronic music often uses synthesizers to evoke the \emph{idea} of an instrument rather than perfectly match the instrument itself (e.g.\ drum machines).
Second, the study of vocal imitation has broad interest in cognitive science, from the emergence of iconicity \citep{hofer2019iconicity, hofer2021learning}, to early language learning in infants \citep{kuhl1996infant, patterson1999matching}, to few-shot learning in birds like parrots, mockingbirds, and songbirds \citep{fiete2007model}. In ongoing work, we are collaborating with a group of cognitive scientists to apply our methods to study such questions about human intelligence.

\begin{acks}
KC was supported by the Hertz Foundation and the NSF Graduate Research Fellowship Program. We would also like to thank Judith Fan and Josh McDermott for early conversations about this project, our online study participants for their time and attention, and the anonymous reviewers for their thoughtful feedback on this paper.
\end{acks}

\bibliography{refs.bib}

\clearpage


\appendix

\end{document}